\documentclass[12pt]{iopart}
\usepackage{graphicx}
\pdfoutput=1
  
\begin{document}

\title[FM Quantum Criticality in the Quasi-1D Heavy-Fermion Metal YbNi$_4$P$_2$]{Ferromagnetic Quantum Criticality in the Quasi-One-Dimensional Heavy Fermion Metal YbNi$_4$P$_2$}

\author{C Krellner, S Lausberg, A Steppke, M Brando, L Pedrero, H Pfau, S Tenc\'e, H Rosner, F Steglich, and C Geibel}
\address{Max Planck Institute for Chemical Physics of Solids, D-01187 Dresden, Germany }
\ead{krellner@cpfs.mpg.de}

\begin{abstract}
We present a new Kondo-lattice system, YbNi$_4$P$_2$, which is a clean heavy-fermion metal with a severely reduced ferromagnetic ordering temperature  at $T_C=0.17$\,K, evidenced by distinct anomalies in susceptibility, specific-heat, and resistivity measurements. The ferromagnetic nature of the transition, with only a small ordered moment of $\sim 0.05\,\mu_B$, is established by a diverging susceptibility at $T_C$ with huge absolute values in the ferromagnetically ordered state, severely reduced by small magnetic fields. Furthermore, YbNi$_4$P$_2$ is a stoichiometric system with a quasi-one-dimensional crystal and electronic structure and strong correlation effects which dominate the low temperature properties. This is reflected by a stronger-than-logarithmically diverging Sommerfeld coefficient and a linear-in-$T$ resistivity above $T_C$ which cannot be explained by any current theoretical predictions. These exciting characteristics are unique among all correlated electron systems and makes this an interesting material for further in-depth investigations.
\end{abstract}

\maketitle

Phase transitions are one of the most fascinating phenomena and a central topic in solid-state physics. 
While classical phase transitions driven by thermal fluctuations have been extensively studied, the current interest is on continuous quantum phase transitions which occur at zero temperature and are caused by collective quantum fluctuations between competing ground states \cite{NPFocus:2008}. 
Namely, the emergence of intriguing new states of matter at such a quantum critical point (QCP) was explored in great detail in Lanthanide-based heavy-fermion systems \cite{Mathur:1998, Stewart:2001, Park:2006, Friedemann:2009, Stockert:2011}. 
However, despite these studies over more than two decades, no $4f$-based material has been found with a ferromagnetic (FM)-to-paramagnetic quantum phase transition. The reason for this lapse remains hotly debated \cite{Sullow:1999, Kirkpatrick:2003, Yamamoto:2010}. Here, we present a new heavy-fermion metal, YbNi$_4$P$_2$, which indeed presents FM quantum criticality above a low-lying FM transition temperature. 

\begin{figure}
\begin{center} 
\includegraphics[width=0.8\textwidth]{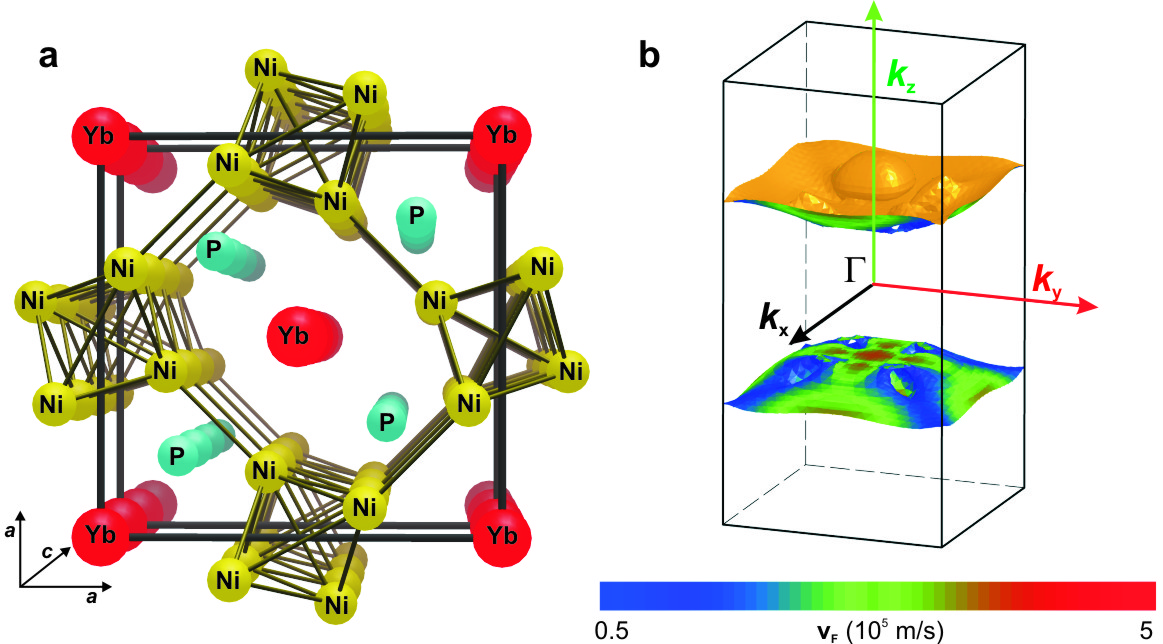}%
\caption{\label{fig1}Quasi-one-dimensional crystal and electronic structure of YbNi$_4$P$_2$.\\
\textbf{a},  Stereoscopic view of the tetragonal crystal structure along $c$ with the Yb chains located in the channels between chains of edge-connected Ni tetrahedra. Details of the structural refinement of our samples are given in the Appendix B.  \\
\textbf{b}, Topology of one of the uncorrelated Fermi surfaces with the most pronounced one-dimensional character manifested in two nearly flat sheets, well separated along $k_z$. Note that the viewing direction is rotated by 90$^{\circ}$ compared to the crystal structure. The size of the Fermi velocity, $v_F$, of the mainly Ni-$3d$ states dominating the density of states at $E_F$ is colour-coded.}
\end{center}
\end{figure}

Intermetallic pnictides have recently become a focus in the solid state physics community as a result of the discovery of high-temperature superconductivity in the RFeAsO (R = rare earth) and AFe$_2$As$_2$ (A = alkali metal) systems with different substitutions (for a review see e.g. Ref.~\cite{Paglione:2010}). While these compounds present a pronounced quasi-two-dimensional character, the pnictides crystallising in the ZrFe$_4$Si$_2$ structure type are quasi-one-dimensional and have been poorly investigated \cite{Chikhrij:1986, Pivan:1989, Jeitschko:1990}. In Fig.~\ref{fig1}a we present the tetragonal ZrFe$_4$Si$_2$ structure type ($P4_2/mnm$) of YbNi$_4$P$_2$ which can be viewed as isolated chains (along the $c$ direction) of edge-connected Ni tetrahedra, with adjacent chains linked by Ni-Ni bonds between corners of the tetrahedra. The Yb atoms are located in the channels between these Ni tetrahedral chains. The quasi-one-dimensional character in the Yb and in the  Ni network, as well as the geometrical frustration between neighboring  Yb chains which are shifted by $c/2$, are prone to cause strong quantum fluctuations if Yb is magnetic (Yb$^{3+}$). 

Previously, only a susceptibility curve at high temperatures of an impure YbNi$_4$P$_2$ sample was reported by Deputier \textit{et al.} \cite{Deputier:1997}. These authors suggested magnetic Yb$^{3+}$ ions and a non-magnetic contribution of the ``Ni$_4$P$_2$'' sublattice revealed by weak Pauli paramagnetism in the reference compound YNi$_4$P$_2$, in contrast to the first measurements by Chikhrij \textit{et al.} \cite{Chikhrij:1991} which suggest a ferromagnetically ordered sublattice above room temperature, most probably due to Ni impurity phases. We succeeded in preparing single-phase YbNi$_4$P$_2$ polycrystals. Our susceptibility measurements, $\chi(T)$, indicate a clear Curie-Weiss behaviour between 50 and 400\,K with a Weiss temperature, $\Theta_W=35$\,K, and an effective moment, $\mu_{\textrm{eff}} = 4.52\,\mu_B$, due to magnetic Yb$^{3+}$ ions, thus supporting the absence of a Ni moment. The lattice parameters $a=7.0565(2)$\,\AA\, and $c=3.5877(1)$\,\AA\, refined by simple least squares fitting agree well with the reported structure data \cite{Kuzma:2000}. A complete structure refinement based on tiny single crystals is given in the Appendix B.

To gain insight into the electronic structure of YbNi$_4$P$_2$, we performed band structure calculations of the ``Ni$_4$P$_2$'' sublattice. For this purpose, the 4$f^{13}$ electrons are frozen as core states and are not allowed to hybridise with the conduction electrons as the strong Coulomb correlations of the Yb $4f$ electrons would give incorrect results in the mean-field approximation. Two main results can be inferred from the uncorrelated band structure: First, the three main Fermi surfaces have a predominantly one-dimensional (1D) character: The most prominent one is visualised in Fig.~\ref{fig1}b with two nearly flat sheets (well separated along $k_z$), which is typical for a 1D system in real space. Therefore, not only the crystal structure, but also the electronic structure suggest that YbNi$_4$P$_2$ is a quasi-1D system, unique among Kondo-lattice systems. Second, spin-polarised calculations clearly demonstrate the absence of Ni-related magnetism in YbNi$_4$P$_2$, although the main contributions to the density of states at the Fermi energy, $E_F$, result from Ni-$3d$ states.

\begin{figure}
\begin{center}
\includegraphics[width=0.8\textwidth]{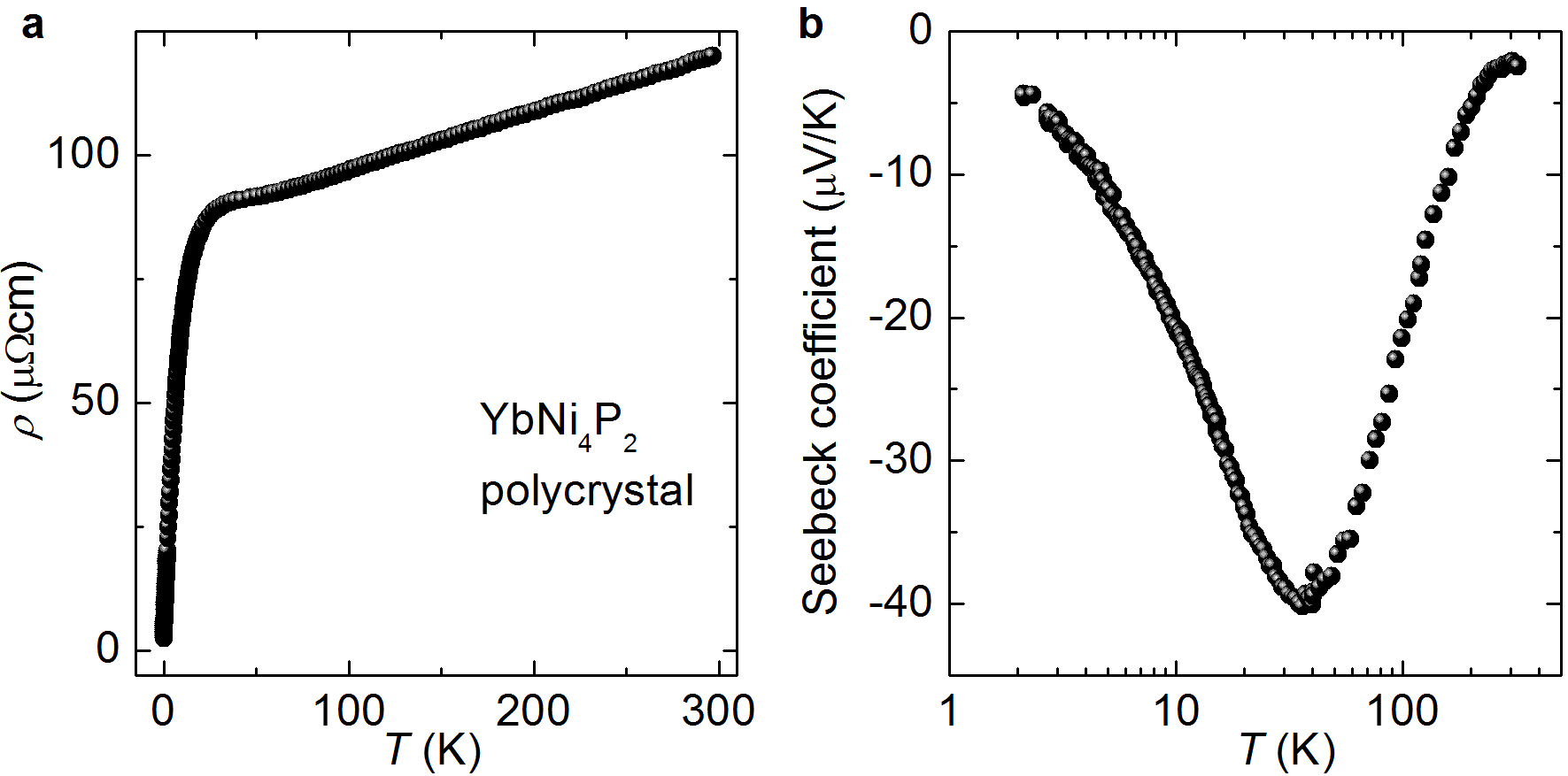}%
\caption{\label{fig2}YbNi$_4$P$_2$: A metallic Kondo lattice.\\
\textbf{a}, The temperature dependence of the resistivity is overall metallic with a pronounced drop below 30\,K, marking the onset of coherent Kondo scattering. The high residual resistivity ratio, $\rho_{300K}/\rho_0= 50$, demonstrates that our polycrystalline sample is very clean.\\
\textbf{b}, The temperature dependence of the Seebeck coefficient shows a distinct minimum at 35\,K, characteristic for Yb-based Kondo lattices. The large absolute values near this minimum are commonly ascribed to a strongly energy-dependent quasiparticle density of states at $E_F$.}
\end{center}
\end{figure}

The temperature dependence of the resistivity, $\rho(T)$, and the Seebeck coefficient, $S(T)$, between 2 and 300\,K (Fig.~\ref{fig2}) evidence YbNi$_4$P$_2$ as a Kondo lattice with strong interactions between $4f$ and conduction electrons. $\rho(T)$ decreases linearly down to 50\,K with a room temperature value of $\rho_{300K}\cong 120\,\mu\Omega$cm, before dropping rapidly below 20\,K due to the onset of coherent Kondo scattering. The correspondingly large residual resistivity ratio, $\rho_{300K}/\rho_0=50$, reveals a long electronic mean free path at low $T$, which demonstrates the high quality of our polycrystalline sample.  Assuming that the linear decrease down to 50\,K dominantly results from phonon scattering, the magnetic part of the resistivity reveals a typical Kondo maximum around 30\,K. The presence of strong hybridisation between the $4f$ and the conduction electrons in YbNi$_4$P$_2$ is further supported by thermopower measurements (see Fig.~\ref{fig2}b). $S(T)$ is negative within the whole temperature range investigated, a fact that is well established in Yb-based Kondo lattices. Moreover, $S(T)$ presents a pronounced minimum at 35\,K with absolute values as high as 40\,$\mu$V/K. Extrema in $\rho(T)$ and $S(T)$ originate from Kondo scattering on the ground and the excited crystal electric field (CEF) levels \cite{Kohler:2008}. Since the CEF scheme of YbNi$_4$P$_2$ is presently unknown, a reliable estimate of the Kondo energy scale, $T_K$ (for the lowest-lying CEF Kramers doublet), can only be obtained by means of the magnetic entropy calculated from the specific heat data, discussed below. At 4\,K, the entropy reaches $0.5 R\ln 2$, establishing a doublet ground state with $T_K \cong  8$\,K.

\begin{figure}
\begin{center}
\includegraphics[width=0.8\textwidth]{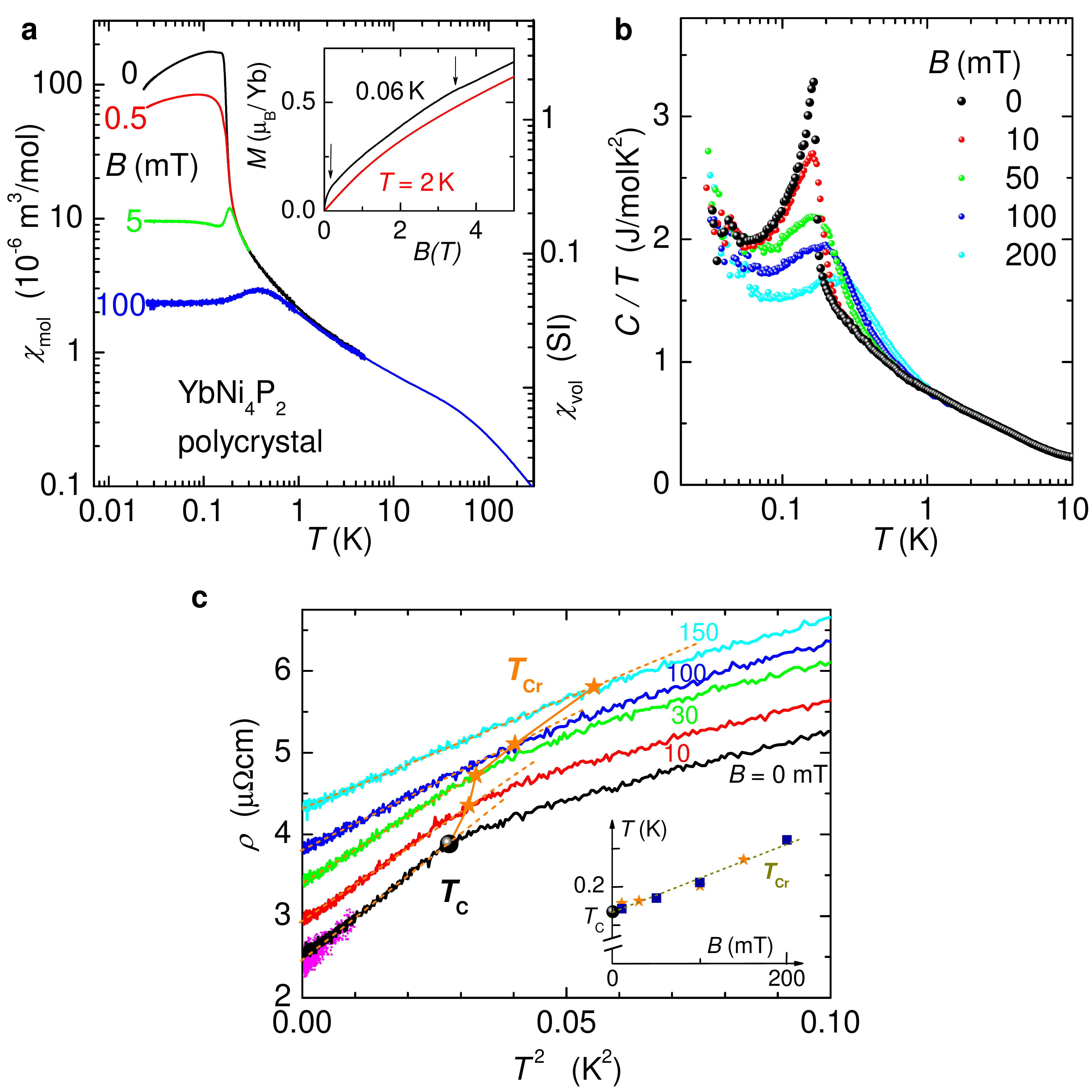}%
\caption{\label{fig3}Low-lying ferromagnetic transition at $T_C=0.17$\,K in YbNi$_4$P$_2$.\\
\textbf{a}, Temperature-dependent a.c. susceptibility, $\chi_{mol}(T)$, at selected magnetic fields reveals a FM phase transition at $T_C$ with very high absolute values. The right hand ordinate is divided by the molar volume, representing the dimensionless volume susceptibility in SI units.
In the inset, the field dependence of the magnetisation below (black line) and above (red line) $T_C$ is shown. The arrow on the left hand side indicates the step-like increase due to the small ordered FM moment, $M_{\textrm{ord}}\cong 0.05\,\mu_B$, obtained by extrapolating the linear $M(H)$ curve between 0.1 and 0.2\,T to zero field. The right hand arrow marks an anomaly at higher field either representing the rotation of the FM moments along the magnetic hard direction or the complete suppression of the Kondo screening.\\
\textbf{b}, Low-temperature specific heat plotted as $C/T$ versus $T$ on a logarithmic scale at zero and low values of the applied magnetic field without subtraction of any hyperfine nor phononic contribution. A sharp $\lambda$-type anomaly at $T_C$ confirms the second-order phase-transition and the high quality of our sample. With increasing field, the peak broadens and transforms into a Schottky-type anomaly with a maximum at $T_{Cr}$. \\
\textbf{c}, Temperature dependence of the resistivity plotted as a function of $T^2$, dashed lines indicate the formation of a Landau-Fermi liquid ground state. For the sake of clarity, the different curves for $B>0$ are shifted by $0.5\,\mu\Omega$cm, respectively. For the zero field curve, we additionally present data measured with very low excitation current but less resolution (magenta curve) to demonstrate the absence of heating effects above 0.07\,K. The zero-field transition at $T_C$ (black symbol) becomes a crossover at $T_{Cr}$ (orange stars) in finite field, shown together in the inset with the specific heat crossover temperatures (blue squares) in a $B$-$T$-phase diagram.}
\end{center}
\end{figure}

We now turn to the exciting low-temperature properties of YbNi$_4$P$_2$ presented in Fig.~\ref{fig3}. Below $T = 10$\,K, the a.c. susceptibility, $\chi$, increases and nearly diverges towards $T_C=0.17$\,K where the $4f$ moments undergo a FM phase transition. The temperature and magnetic field dependence of $\chi$ are distinctive for a ferromagnetically ordered material (see Fig.~\ref{fig3}a). The FM nature of the magnetic phase is one of our key results since the majority of the presently known Kondo lattices order antiferromagnetically. Most prominent are the huge absolute values of $\chi$ in zero magnetic field. To assess the magnitude of the divergence at $T_C$, we scaled the right hand ordinate of Fig.~\ref{fig3}a to the dimensionless volume susceptibility, e.g., $\chi_{vol} = \chi_{mol}/V_{mol}$ with the molar volume, $V_{mol}=53.8 \times  10^{-6}$\,m$^3$/mol. To better compare this with the susceptibility of prototypical ferromagnets as e.g. iron, we need to consider two more facts. First, the ferromagnetically ordered moment in YbNi$_4$P$_2$ is a factor of 20 times smaller (see below) and second, the density of the magnetic Yb ions is 7.5 times lower than in iron. Therefore, the scaled relative permeability for YbNi$_4$P$_2$ is as high as $\sim 500$, very similar to iron. The fact that the magnetic ordering temperature of the Yb moments is so strongly reduced results primarily from the pronounced Kondo interactions. Further evidence for ferromagnetism comes from the very pronounced field dependence of the susceptibility which is suppressed by a factor of two, applying a small field of $B=0.5$\,mT. This makes an AFM or helical order rather unlikely. However, we cannot completely exclude a magnetic ordering vector with a very tiny $q$, but also in such a case the FM interaction would be the most dominant one. For $B\geq 5$\,mT, a broad maximum develops in $\chi(T)$, which shifts to higher temperatures with increasing fields due to the energetic stabilisation of the FM ground state in a magnetic field. In the inset of Fig.~\ref{fig3}a we present the magnetisation data below and above $T_C$ obtained from the same sample, confirming the FM nature of the magnetic transition. At 60\,mK, the magnetisation is strongly nonlinear and increases steeply with field below $B= 150$\,mT due to a small ordered FM moment of $M_{\textrm{ord}}\cong 0.05\,\mu_B$, followed by a second kink at $B=3.5$\,T, which either represents the rotation of the FM moments along the magnetic hard direction (perpendicular to the direction of the FM moment) or the complete suppression of the Kondo screening. Above $T_C$, both anomalies are absent and the polarised moment amounts to $M\cong 0.6\,\mu_B$ at $B=5$\,T. More detailed magnetisation measurements at low fields are presented in the Appendix A. 

Specific heat data, shown in Fig.~\ref{fig3}b, strongly confirm that YbNi$_4$P$_2$ is a magnetically ordered heavy-fermion system. A sharp $\lambda$-type anomaly is observed at $T_C$, establishing a second-order phase transition into the FM phase. Well below $T_C$, a huge Sommerfeld coefficient, $\gamma_0\cong 2000$\,mJ/molK$^2$, reflects the existence of heavy quasiparticles with an electronic mass two to three orders of magnitude bigger than the bare electron mass. Below 50\,mK, the scattering of the data points significantly increases, preventing a more accurate determination of the Sommerfeld coefficient. Remarkably, $C/T$ is larger below $T_C$ than above the FM ordering, indicating strong fluctuations within the ferromagnetically ordered state, similar to the antiferromagnetic (AFM) phase in YbRh$_2$Si$_2$ \cite{Oeschler:2008}. Integrating $C/T$ over temperature reveals an entropy gain associated with the anomaly at $T_C$ of only about $0.02 R\ln2$. This is in accordance with the small value of the ordered moment derived from our magnetisation data and provides evidence for rather weak FM order in YbNi$_4$P$_2$. 

The onset of magnetic order is further established by the freezing out of spin-disorder scattering, i.e., a distinct reduction of the electrical resistivity at $T_C$, presented as $\rho$ vs. $T^2$ in Fig.~\ref{fig3}c. Below $T_C$, the resistivity follows a straight line down to the lowest measured temperatures, i.e. $\rho(T)=\rho_0+AT^2$ with $\rho_0\cong 2.5\,\mu\Omega$cm and $A\cong 52\,\mu\Omega$cm/K$^2$. Therefore, all measured quantities present the hallmarks of a  Landau-Fermi-liquid ground state within the FM ordered phase and can be characterised by the $T\rightarrow 0$ limits of the susceptibility, $\chi_0$, Sommerfeld coefficient of the electronic specific heat, $\gamma_0$, and the resistivity coefficient $A$. The so-called Kadowaki-Woods ratio, $A/\gamma_0^2$, amounts to approximately $13\,\mu\Omega$cmK$^2$mol$^2$/J$^2$, close to the universal value of $10\,\mu\Omega$cmK$^2$mol$^2$/J$^2$ for heavy-fermion compounds \cite{Kadowaki:1986}. The Sommerfeld-Wilson ratio, $W=\pi^2k_B^2/(\mu_0\mu_{\textrm{eff}}^2)\cdot \chi_0/\gamma_0$ with $\mu_{\textrm{eff}}\cong 2\,\mu_B$, is found to be huge compared to all other heavy fermion systems where typically values between 1 and 5 are observed \cite{Fisk:1987}. In YbNi$_4$P$_2$, a giant value of $W$ is observed within the ferromagnetically ordered state ($W_{T\rightarrow 0}\cong  350$) and $W$ is still substantially enhanced well above $T_C$, $W_{T=0.3\,K}\cong  20$. This leads further support to very strong FM quantum fluctuations in YbNi$_4$P$_2$.

The effect of an applied magnetic field is similar for all presented quantities: Already a small ($\cong 10$\,mT) magnetic field extends the Landau-Fermi-liquid ground state to higher temperatures and is  accompanied by a reduction of the effective mass reflected in decreased values of $\gamma_0$, $A$, and $\chi_0$. The field-stabilised FM state is characterised by broad crossover maxima at $T_{Cr}$ in $\chi(T)$ and $C(T)/T$, respectively, as the Yb-moments get polarised in the external field at higher temperatures (see inset of Fig.~\ref{fig3}c). Therefore, the system is tuned away from the QCP, in striking contrast to  AFM systems, where an ordered system may be field-tuned through the QCP \cite{Stewart:2001, Heuser:1998, Budko:2004}. 

\begin{figure}
\begin{center}
\includegraphics[width=0.8\textwidth]{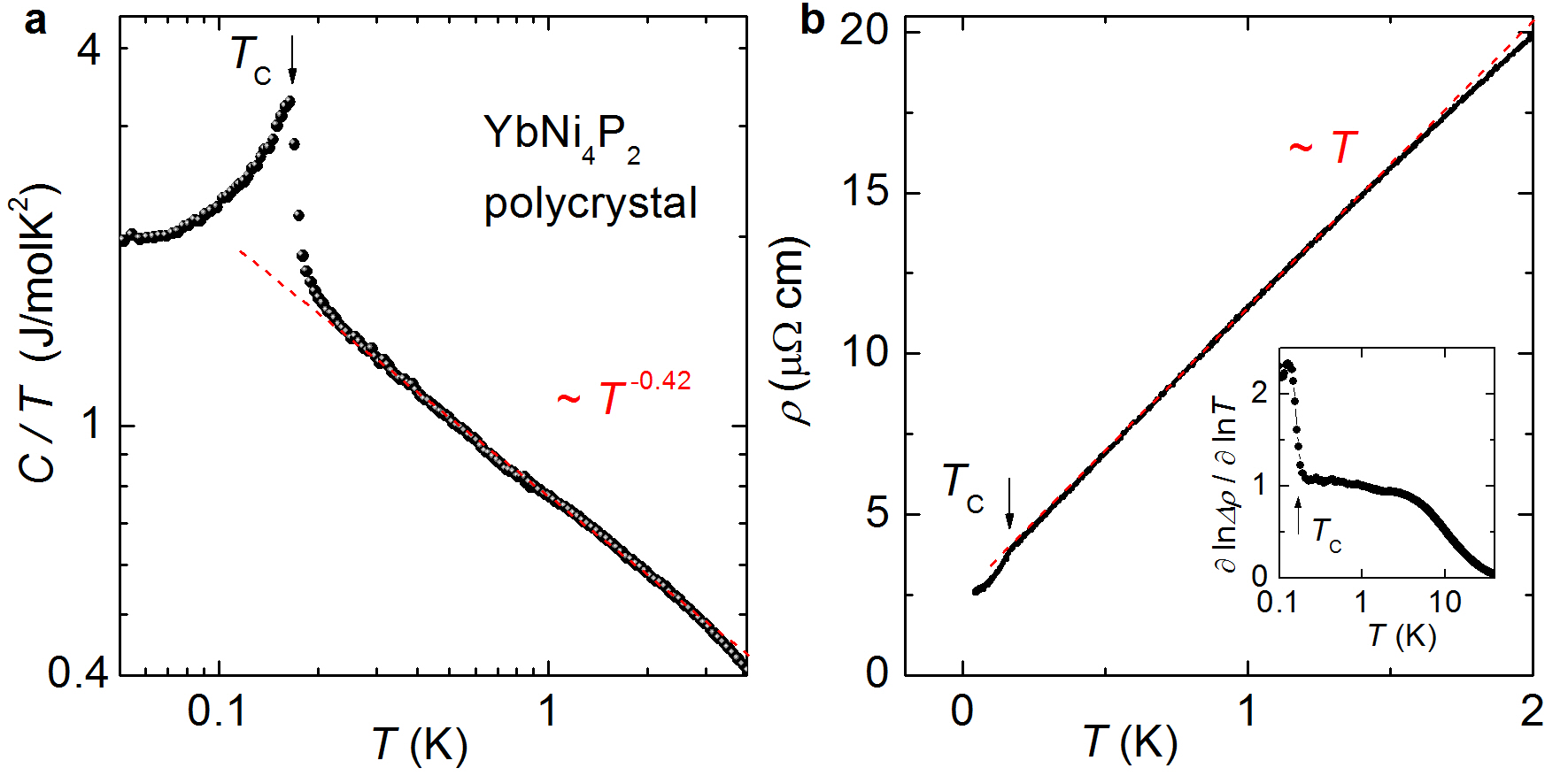}%
\caption{\label{fig4}Pronounced non-Fermi-liquid behaviour above $T_C$ at $B=0$.\\
\textbf{a}, The temperature dependence of the zero-field specific heat divided by temperature in a double logarithmic representation clearly reveals a power law divergence, $C/T\propto T^{-0.42}$, indicated by the red dashed line.\\
\textbf{b}, Linear-in-$T$ resistivity in zero field over more than a decade in temperature. A pronounced drop at $T_C$ precedes a $T^2$ law within the ordered phase. The temperature dependence of the  resistivity exponent, defined as the derivative of $\ln\bigl{[}\Delta \rho(T)\bigr{]}=\ln\bigl{[}\rho(T)-\rho_0\bigr{]}$ with respect to $\ln T$, is presented in the inset.}
\end{center}
\end{figure}

Having established the proximity of YbNi$_4$P$_2$ to a FM QCP we now turn to the quantum critical regime and address the distinct deviations from the predictions of Landau-Fermi-liquid (LFL) theory above $T_C$. They are most pronounced in zero magnetic field, where YbNi$_4$P$_2$ is closest to the QCP. Remarkably, the specific-heat coefficient diverges stronger than logarithmically below $T=3$\,K with a power law $C/T\propto T^{-0.42}$ down to 0.2\,K. At this temperature the classical fluctuations of the FM order parameter set in (see Fig.~\ref{fig4}a). Above $T=1$\,K, there might be a crossover from a power-law to a logarithmic divergence up to $T=7$\,K, better seen in the semi-logarithmic plot of $C/T$ vs. $T$ in Fig.~\ref{fig3}b. The resistivity follows a linear-in-$T$ dependence for $T>T_C$, presented in Fig.~\ref{fig4}b, with a tendency to sub-linear behaviour above $T=2$\,K. For a more quantitative analysis of the resistivity exponent, we consider the logarithmic derivative, $\partial\ln(\rho-\rho_0) / \partial\ln T$. As displayed in the inset of Fig.~\ref{fig4}b, $\partial\ln(\rho-\rho_0) / \partial\ln T$ remains within the value $1.0\pm0.1$ over more than one decade in temperature above $T_C$ before dropping above 5\,K. 

These strong deviations from LFL behaviour along with the low-lying FM ordering  have very general implications for our current understanding of quantum phase transitions. While numerous AFM Kondo systems have been tuned towards a QCP by variation of pressure, doping or magnetic field, appropriate candidates for the study of FM QCPs are extremely rare \cite{Stewart:2001}. Starting deep inside the localised moment regime in Ce-based ferromagnets, the increase of the Kondo interaction with pressure typically leads to an AFM ground state before the QCP is reached \cite{Sullow:1999}. In disordered FM $4f$ systems a peculiar Kondo-cluster-glass state was found, preventing the study of FM quantum criticality \cite{Westerkamp:2009}. On the other hand, in clean itinerant FM $3d$ systems, it is established that no QCP exists, rather the FM transition always ends at a classical critical point (at finite $T$) where a first-order phase transition occurs \cite{Kirkpatrick:2003, Chubukov:2004}. However, disorder can lead to weak signatures of FM quantum criticality as observed in various $3d$ alloy series \cite{Nicklas:1999, Sokolov:2006, Jia:2011}, which theoretically could be explained by disorder-induced rounding effects \cite{Vojta:2003}. Therefore, YbNi$_4$P$_2$ represents the first clean example of FM quantum criticality above a low-lying FM transition, permitting the examination of existing and the development of new theoretical predictions for the temperature dependencies of the relevant physical parameters at a FM QCP. Most theoretical work has been done in systems where the vanishing magnetism can be described within the framework of itinerant spin fluctuation theories \cite{Stewart:2001}. However, our observed power laws in $C(T)/T$ and $\rho(T)$  above $T_C$ deviate strongly from these theoretical descriptions. Presently, we cannot distinguish whether these deviations result from the quasi-one dimensionality of YbNi$_4$P$_2$, for which no calculations have been performed, or whether YbNi$_4$P$_2$ is situated close to a local QCP, where the FM transition is accompanied by a localisation-delocalisation transition of the $f$ electrons. The local QCP scenario is well established for YbRh$_2$Si$_2$ \cite{Friedemann:2010} which presents AFM order below $T_N=72$\,mK, but nonetheless shows strong FM fluctuations in wide parts of its phase diagrams \cite{Gegenwart:2005, Klingner:2011}. We note that in YbRh$_2$Si$_2$, $C(T)/T$ also follows a stronger-than-logarithmically divergence below 300\,mK and a linear-in-$T$ resistivity \cite{Custers:2003}. Furthermore, a sharp $\lambda$-type transition into the antiferromagnetically ordered phase and a strongly enhanced Sommerfeld coefficient below $T_N$ were similarly observed for YbRh$_2$Si$_2$ \cite{Krellner:2009}. Future work has to be undertaken to tune YbNi$_4$P$_2$ towards the non-magnetic side of the QCP. Negative chemical pressure achieved by As substitution on the P site might be the most promising route as the stoichiometric YbNi$_4$As$_2$ forms in the same crystal structure but with non-magnetic Yb$^{2+}$ ions \cite{Deputier:1997} due to the much larger volume of the unit cell.

To conclude, we have discovered a new heavy-fermion Kondo-lattice system with several features that are unique among strongly correlated $4f$ systems. First, both the crystal and the electronic structure of YbNi$_4$P$_2$ are quasi-one-dimensional, originating in distinct quantum fluctuations at low temperatures. Second, YbNi$_4$P$_2$  undergoes a well-defined ferromagnetic phase transition of second order at $T_C=0.17$\,K, growing out of a strongly correlated Kramers doublet ground state with a Kondo temperature, $T_K\cong 8$\,K. Due to the dominant Kondo screening only a tiny ferromagnetically ordered moment ($M_{\textrm{ord}}\cong 0.05\,\mu_B$) with a substantially reduced $T_C$ is observed. The ferromagnetism is evidenced by a diverging susceptibility at $T_C$ with huge absolute values comparable to established ferromagnets. The $4f$ spins remain strongly fluctuating down to the lowest measured temperatures, leading to a large Sommerfeld coefficient, $\gamma_0\cong 2000$\,mJ/molK$^2$, within the ordered phase. Above $T_C$, a remarkable power-law divergence of the specific heat, $C/T\propto T^{-0.42}$, and a linear-in-$T$ resistivity are observed over more than a decade in temperature. Therefore, YbNi$_4$P$_2$ is the first clean system situated in the close vicinity of a ferromagnetic quantum critical point with the lowest lying $T_C$ ever observed among correlated systems. These ferromagnetic quantum fluctuations which dominate the thermodynamic and transport properties well above $T_C$ cannot be explained within any current theoretical framework. Furthermore, YbNi$_4$P$_2$ is a highly stoichiometric system and has the potential to become the prototype of a ferromagnetic quantum critical material.

\section*{Acknowledgements}
We sincerely thank Ralf Weise, N. Caroca-Canales, and C. Bergmann for assistance in sample preparation. Valuable discussions with 
M.~Baenitz,
S.~Friedemann,
A.~Haase,
A.~Jesche,
S.~Kirchner,
H.-H.~Klau\ss, 
R.~K\"uchler,
N.~Mufti,
M.~Nicklas,
R.~Sarkar,
J.~Spehling, 
O.~Stockert,
U.~Stockert, and
S.~Wirth
are gratefully acknowledged.
This work is partially supported through DFG Research Unit 960 and the SPP 1458.

\appendix
\setcounter{section}{0}

\section{Methods}
The polycrystalline samples were prepared by heating a stoichiometric amount of the elements in evacuated quartz ampoules at 1000$^{\circ}$C for 24 hrs. The sample was characterised by powder and single crystalline X-ray  diffraction experiments presented in detail in the Appendix B. 
Electronic band structure calculations were done with the full-potential nonorthogonal local-orbital minimum basis scheme FPLO code (version: FPLO 9.00-31) \cite{Koepernik:1999}. Within the local spin density approximation, the exchange and correlation potential of Perdew and Wang has been used \cite{Perdew:1992}.
The high-temperature measurements of $\rho(T)$, $S(T)$ and $C(T)$, were performed using a commercial Quantum Design PPMS equipped with a $^3$He option. The magnetisation and d.c. susceptibility measurements above 2\,K were conducted using a Quantum Design SQUID VSM. 
All low-temperature measurements were carried out in  $^3$He/$^4$He dilution refrigerators.
The a.c. susceptibility was determined at low frequencies with a modulation field amplitude of 15\,$\mu$T down to 0.02\,K, measured in selected static magnetic fields. Great care was taken to  account for remanent fields. Above 2\,K, the a.c. susceptibility equals the d.c. susceptibility determined by the SQUID VSM.
The specific heat between 0.03 and 1\,K was measured utilising a semiadiabatic heat-pulse method with the sample mounted on a silver platform.
The electrical resistivity $\rho(T)$ was monitored by a standard four-point lock-in technique at 16.67\,Hz down to 0.02\,K.
\begin{figure}[t]
\begin{center}
\includegraphics[width=0.8\textwidth]{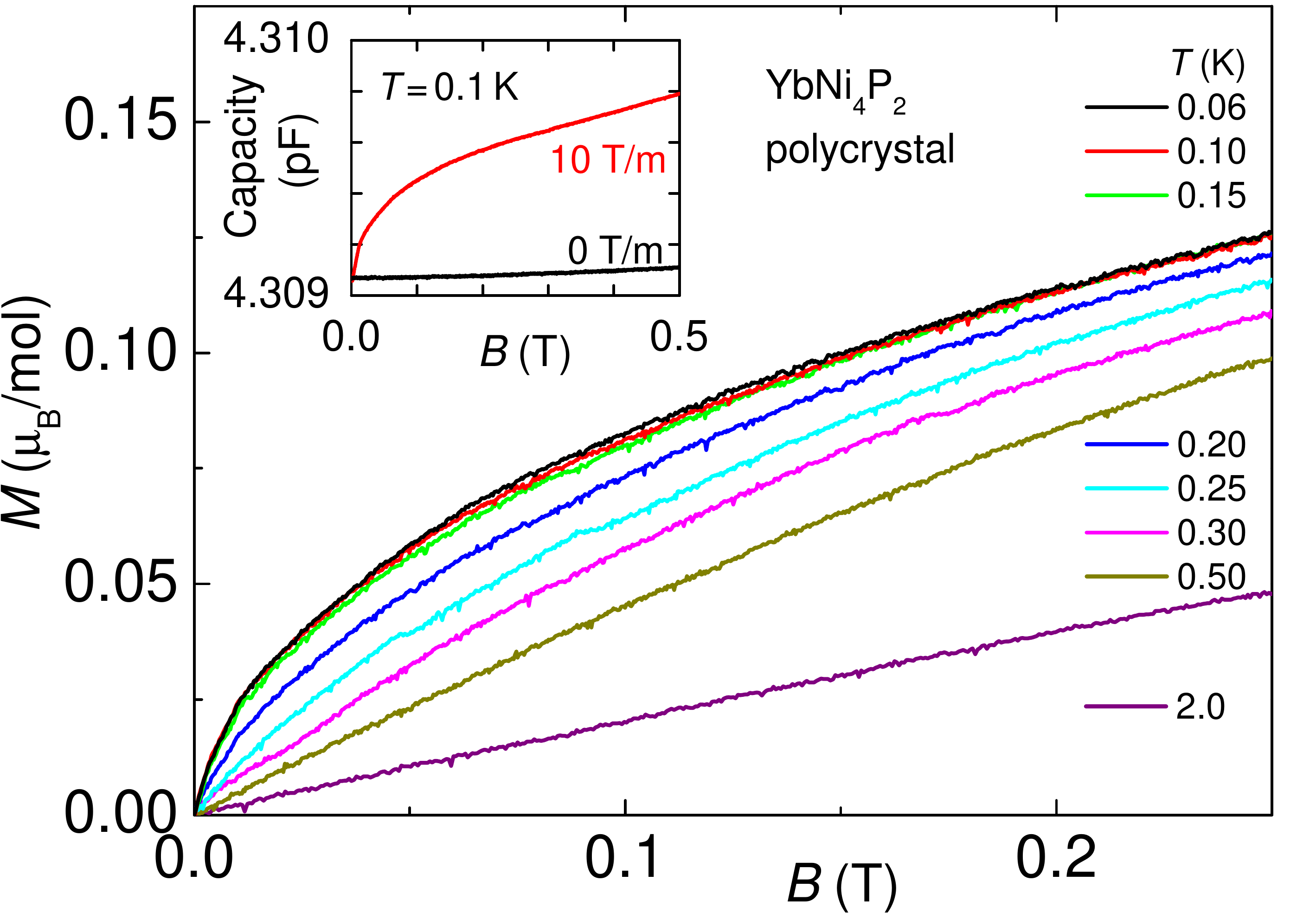}%
\caption{\label{s1}Detailed magnetisation measurements of YbNi$_4$P$_2$ at low magnetic fields for various temperatures below and above $T_C$. The measured capacitance at $T=0.1$\,K with an applied magnetic field gradient (d$B/$d$z=10$\,T/m, red line) is shown in the inset together with the data measured without the field gradient (black line). The latter serves as a background signal and was subtracted to obtain the presented $M(B)$ curves.}
\end{center}
\end{figure}

The magnetic field dependence of the magnetisation $M(B)$ was isothermally measured in a high-resolution Faraday magnetometer down to 0.06\,K \cite{Sakakibara:1994}. Background contributions from the sample platform and the torque exerted on the sample have been subtracted and were determined by measuring the capacitance without a magnetic field gradient (black line in the inset of Fig.~A1). However, the signal of the sample (red line in the inset of Fig.~A1) dominates the field dependence of the capacitance. In the main part of Fig.~A1, more detailed $M(B)$ curves are presented in addition to the data in the inset of Fig.~\ref{fig3}a. Below $T_C$, the $M(B)$ curves are nearly temperature independent but decrease strongly above $T_C$ as expected for a FM system.

\section{X-ray diffraction}

In Fig.~B1, we present the X-ray powder diffraction data of a polycrystalline YbNi$_4$P$_2$ sample recorded on an Imaging Plate Guinier Camera (Huber G670, CuK$\alpha$ radiation, $\lambda=1.5406$\,\AA).  All peaks could be indexed using the ZrFe$_4$Si$_2$ structure type (green lines in Fig.~B1) confirming the absence of any foreign phase in our sample. Unit cell parameters were refined by a least-squares procedure (program package WinCSD \cite{Akselrud:1993}) using the peak positions extracted from powder patterns measured with LaB$_6$ as an internal standard (not shown). 

Furthermore, a complete structural refinement of a small single crystal was carried out on a Rigaku AFC7 diffractometer equipped with a Saturn 724+ CCD detector using MoK$\alpha$ radiation. The crystallographic information and experimental details of this refinement are presented in Table B1 and were done using the program package ShelX \cite{Sheldrick:2008}. The structure is in  agreement with the literature \cite{Kuzma:2000}, however, our refinement based on single crystals is more precise than the reported powder data.

\begin{figure}[t]
\begin{center}
\includegraphics[width=0.8\textwidth]{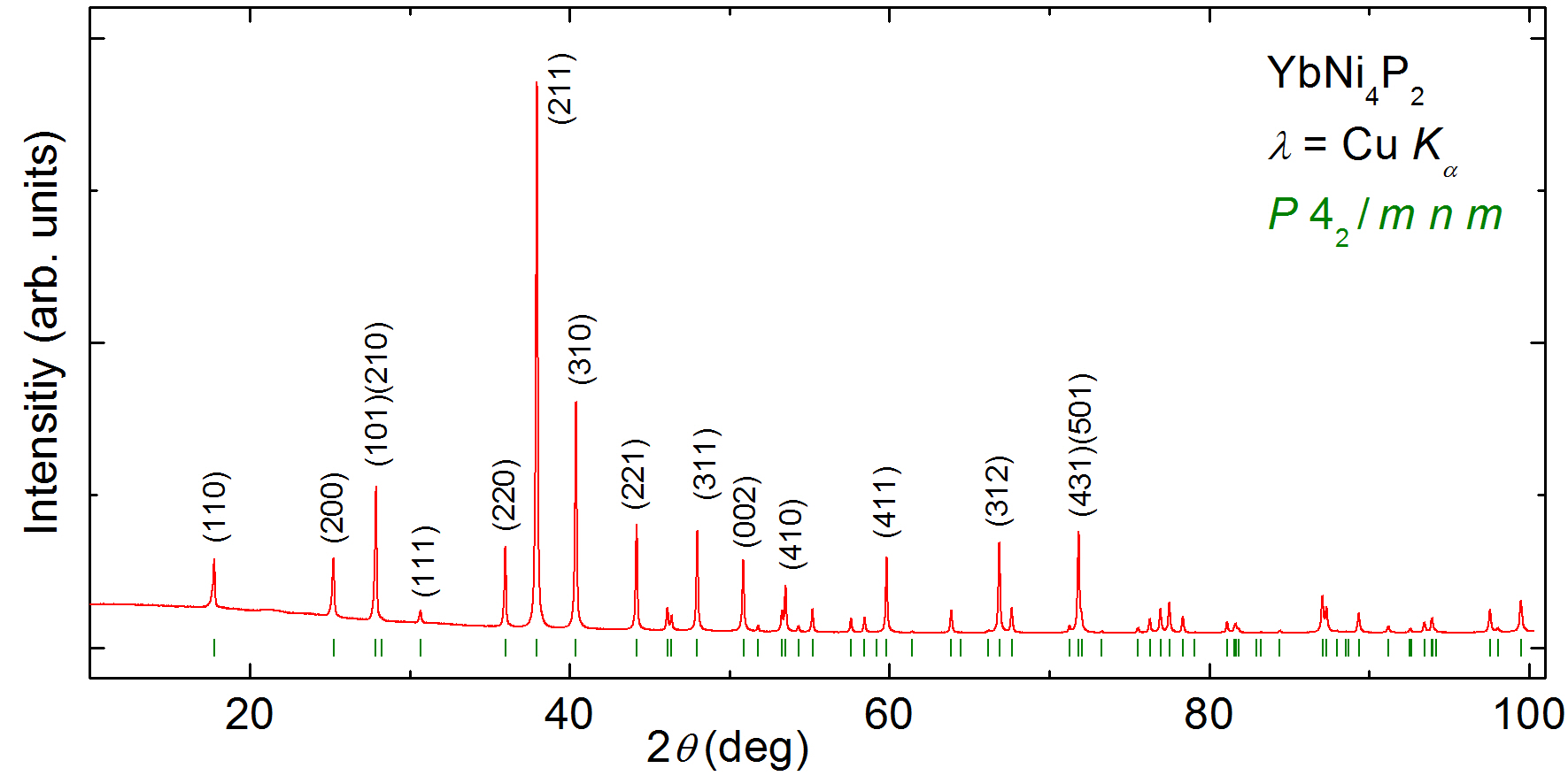}%
\caption{\label{s2}Powder diffraction pattern of YbNi$_4$P$_2$ at room temperature. Green lines denote the peak positions for the $P4_2/m n m$ structure type.}
\end{center}
\end{figure}

These crystallographic data and the refinement corroborate that YbNi$_4$P$_2$ is structurally highly ordered without significant crystallographic defects. For the study of quantum critical phenomena at low temperature this is an important conclusion because disorder can often prevail the intrinsic properties particular for materials where the dominating energy scales are small. Furthermore, the excellent stoichiometric quality of our YbNi$_4$P$_2$ samples is in nice agreement with the presented physical properties at low temperature, e.g., sharp anomalies at $T_C=0.17$\,K and a large residual resistivity ratio.

\begin{table}[h]
\begin{center}
\caption{\label{TabCryst} Crystallographic data and experimental details for the single crystal refinement of YbNi$_4$P$_2$ at room temperature.\\
Crystallographic data: space group $P4_2/m n m$ (no. 136), $a = 7.0565(2)$\,\AA, $c = 3.5877(1)$\,\AA, $V = 178.65(1)$\,\AA$^3$ (lattice parameters were obtained from powder diffraction data), $Z = 2$, $F(000) = 424$, $U_{13}=U_{23}=0$, $\rho_{\textrm{calc}} = 8.73$\,g/cm$^3$, $\mu(\textrm{MoK}_{\alpha}) = 47.24$\,mm$^{-1}$, crystal dimensions $20\times 35 \times 65\,\mu$m$^3$.\\ 
Collected data: Mo K$\alpha$ radiation ($\lambda = 0.71073$\,\AA), $2\theta_{\textrm{max}}=70^{\circ}$, 1172 measured reflections, 238 symmetry independent reflections, 226 observed reflections [$I>2\sigma(I)$], $R_{\textrm{int}} = 0.040$, 15 parameters, $R(F) = 0.027$, $wR(F^2)=0.059$, GooF$=1.12$, numerical absorption $T_{\textrm{min}}/T_{\textrm{max}} = 0.3572$, $\delta\rho_{\textrm{min (max)}}= -3.11\,(2.13)$\,e/\AA$^3$.}
\footnotesize
\begin{tabular}{|l|c|c|c|c|c|c|c|c|c|}
\hline 
At.	&	Site & $x/a$ &	$y/a$  & $z/c$ &   $U_{11}$ &   $U_{22}$ &	  $U_{33}$ &  $U_{12}$ & $U_{eq}$ \\ \hline   
			
Yb &	2b	& 0 & 0 & 0.5  & 0.0070(2) & $U_{11}$ & 0.0077(3) & -0.0002(1) & 0.0072(2)\\	
Ni &	8i	& 0.3359(1) & 0.0840(1) & 0  & 0.0086(4) & 0.0077(3) & 0.0083(4) & -0.0008(2) & 0.0082(2)\\	
P  &	4g	& 0.2195(2) & $-x/a$ & 0  & 0.0083(5) & $U_{11}$ & 0.0076(7) & -0.0004(6) & 0.0081(4)\\	 \hline \hline 
\end{tabular}
\end{center}
\end{table}

\end{document}